\documentclass[a4paper,11pt]{article}
\usepackage{amsmath,amssymb,amsfonts,amsthm,latexsym}

\usepackage{pos}
\usepackage{graphicx}
\usepackage{tabularx}
\usepackage{multicol}
\usepackage{hyphenat}

\usepackage{slashed}
\usepackage{units}
\usepackage{siunitx}
\usepackage{subcaption}
\usepackage{lineno}

\title{Chromoelectric and chromomagnetic correlators at high temperature from gradient flow}
\ShortTitle{Chromoelectric and chromomagnetic correlators from gradient flow}

\author*[a,1]{Julian Mayer-Steudte}
\author[a,b,c,1]{Nora Brambilla}
\author[a,1]{Viljami Leino}
\author[d,1]{Peter Petreczky}
\note{TUMQCD collaboration}

\affiliation[a]{Physik Department, Technische Universit\"at M\"unchen,\\
James-Franck-Strasse 1, 85748 Garching, Germany}

\affiliation[b]{Institute for Advanced Study, Technische Universit\"at M\"unchen,\\
Lichtenbergstrasse 2 a, 85748 Garching, Germany}

\affiliation[c]{Munich Data Science Institute, Technische Universit\"at M\"unchen, \\
Walther-von-Dyck-Strasse 10, 85748 Garching, Germany}

\affiliation[d]{Physics Department, Brookhaven National Laboratory,\\ Upton, New York 11973, USA}

\emailAdd{julian.mayer-steudte@tum.de}
\emailAdd{nora.brambilla@ph.tum.de}
\emailAdd{viljami.leino@tum.de}
\emailAdd{petreczk@bnl.gov}

\abstract{The heavy quark diffusion coefficient is encoded in the spectral functions of the chromoelectric and the chromomagnetic correlators that are calculable on the lattice. We study the chromoelectric and the chromomagnetic correlator in the deconfined phase of SU(3) gauge theory using Symanzik flow at two temperatures $1.5T_c$ and $10000 T_c$, with $T_c$ being the phase transition temperature. To control the lattice discretization errors and perform the continuum limit we use several temporal lattice extents $N_t=16,20,24$ and 28. We observe that the flow time dependence of the chromomagnetic correlator is quite different from chromoelectric correlator most likely due to the anomalous dimension of the former as has been pointed out recently in the literature.}

\FullConference{%
 The 38th International Symposium on Lattice Field Theory, LATTICE2021
  26th-30th July, 2021
  Zoom/Gather@Massachusetts Institute of Technology
}

\begin{document}
\maketitle

\section{Introduction}
The quark gluon plasma is an extreme state of matter at high temperatures, which can be probed in relativistic heavy ion collisions \cite{Pasechnik:2016wkt}. Measurements in heavy ion collisions are used to test the QCD phase diagram, which also gives important implications for cosmology and neutron stars. The interpretation of experimental data requires first principle QCD calculations. The heavy quark momentum diffusion coefficient is one of the quantities of interest and can be related to experimental observables such as nuclear suppression 
factor and elliptic flow of open heavy flavor hadrons. However, perturbative calculations of the heavy quark diffusion at LO \cite{Moore:2004tg,Svetitsky:1987gq} and at NLO \cite{Caron-Huot:2008dyw} lead to quite different
results, suggesting that these calculations are not
reliable. Hence, non-perturbative calculations are needed.

The heavy quark diffusion coefficient has been defined through the correlator of two chromoelectric fields \cite{Casalderrey-Solana:2006fio}. Using nonrelativistic effective field theories and open quantum systems, the same object
emerges in the description of quarkonium evolution in medium \cite{Brambilla:2017zei}. Previous lattice calculations of this object \cite{Meyer:2010tt, Francis:2011gc, Banerjee:2011ra,Francis:2015daa,Brambilla:2020siz} were performed in quenched QCD and the heavy quark diffusion constant is obtained as the intercept of the chromolectric correlator spectral function at zero frequency \cite{Caron-Huot:2008dyw}.
The chromoelectric correlator is very noisy and to reduce the noise
the multi-level algorithm has been used in the above studies.
Recently, it has been suggested to use the gradient flow as an alternative
noise reduction method and this approach turned out to be successful in
calculating the heavy quark diffusion coefficient at $1.5T_c$
\cite{Altenkort:2020fgs}. Furthermore,
in Ref. \cite{Bouttefeux:2020ycy} the $1/M$ effects in heavy quark
diffusion were considered, with $M$ being the heavy quark mass. These effects
are encoded in the chromomagnetic correlator \cite{Bouttefeux:2020ycy}.
In this contribution we will present numerical lattice calculations of the chromoelectric and chromomagnetic correlators 
at temperatures of $1.5T_c$ and $10^4T_c$. We will use the gradient flow method for
noise reduction which also ensures the non-perturbative renormalization.

\section{Physical Background}\label{sec:physical_background}
Within a quark gluon plasma heavy quarks can diffuse through the medium. Thereby, the heavy quark momentum is changed by random kicks from the medium which can be described by a Brownian motion encoded into the Langevin dynamics
\begin{linenomath}\begin{align}
    \dot{\mathbf{p}}-\eta \mathbf{p}&=\mathbf{f}(t) ,
\end{align}\end{linenomath}
where $\dot{\mathbf{p}}$ is the temporal derivative of the heavy quark momentum $\mathbf{p}$, $\eta$ is the drag coefficient, and $\mathbf{f}$ is the random force acting on the heavy quark that satisfies
\begin{linenomath}\begin{align}
    \langle f_i(t)\rangle &= 0,\ \langle f_i(t')f_j(t)\rangle = \kappa \delta_{ij}\delta(t-t') ,
\end{align}\end{linenomath}
where $\kappa$ is the heavy quark momentum diffusion coefficient. The drag coefficient and the heavy quark momentum diffusion coefficient are connected through the 
Einstein equation, which after including the leading relativistic correction
reads \cite{Bouttefeux:2020ycy}
\begin{linenomath}\begin{align}
    \eta \approx \frac{\kappa}{2MT}\left( 1-\frac{5T}{2M}\right),
\end{align}\end{linenomath}
where $T$ is the temperature and $M$ the mass of the heavy quark.

As mentioned above, the heavy quark diffusion coefficient can be obtained from
the Euclidean chromoelectric correlator \cite{Casalderrey-Solana:2006fio, Caron-Huot:2008dyw} and the chromomagnetic correlator \cite{Bouttefeux:2020ycy} 
\begin{linenomath}\begin{align}
    G_E(\tau)&=-\frac{1}{3}\frac{\sum_i \mathrm{Tr}\langle U(\beta;\tau)gE_i(\tau)U(\tau;0)gE_i(0) \rangle}{\mathrm{Tr}\langle U(\beta;0)\rangle}\label{eq:chromo_electric_correlator}\\
    G_B(\tau)&=\frac{1}{3}\frac{\sum_i \mathrm{Tr}\langle U(\beta;\tau)gB_i(\tau)U(\tau;0)gB_i(0) \rangle}{3\mathrm{Tr}\langle U(\beta;0)\rangle}\label{eq:chromo_magnetic_correlator},
\end{align}\end{linenomath}
where $\tau$ is the temporal separation of the field insertions and
$E_i$ and $B_i$ are the chromoelectric and chromomagnetic field respectively.
The Wilson lines $U(\tau;0)$ in the above
expression are in the fundamental representation and wrap
around the periodic boundary making the correlators gauge invariant.
The denominator cancels out the self energy divergence in the above
expression. The chromoelectric correlator defined above also plays an important role in quarkonium suppression \cite{Brambilla:2017zei}.

The chromoelectric and chromomagnetic
correlators are connected to the respective diffusion coefficient $\kappa_{E/B}$ 
through the spectral function
\begin{linenomath}\begin{align}
    G_{E/B}(\tau)=\int _0^\infty \frac{\mathrm{d}\omega}{\pi}\rho_{E/B}(\omega)\frac{\cosh \left( \frac{\beta}{2}-\tau\right) \omega}{\sinh \frac{\beta\omega}{2}},\ \ \ \
    \kappa_{E/B}=\lim _{\omega\rightarrow 0}\frac{2T\rho_{E/B}(\omega)}{\omega}.
\end{align}\end{linenomath}
The spectral functions $\rho_{E/B}$ are known perturbatively, see Ref. \cite{Burnier:2010rp} for $\rho_E$ at NLO and \cite{Bouttefeux:2020ycy} for $\rho_B$ at LO. The total heavy quark momentum diffusion coefficient results into
\begin{linenomath}\begin{align}
    \kappa \approx \kappa_{E} + \frac{2}{3}\langle \mathbf{v}^2\rangle\kappa _B,
\end{align}\end{linenomath}
where $\langle \mathbf{v}^2\rangle \sim T/M$ is the velocity of the heavy quark \cite{Bouttefeux:2020ycy}. Therefore, the chromoelectric correlator contributes in the leading order of the heavy quark mass and the chromomagnetic correlator in the first order correction, since $\langle \mathbf{v}^2\rangle$ is of order $T/M$ \cite{Bouttefeux:2020ycy}. In Ref. \cite{Burnier:2010rp} $\kappa _E$ is determined up to LO and in Ref. \cite{Bouttefeux:2020ycy} $\kappa _B$, where the LO of $\kappa _E$ starts at $\mathcal{O}(g^2)$ and the LO of $\kappa _B$ at $\mathcal{O}(g^4)$.

\section{Implementation and results}\label{sec:impl_res}

On the lattice the chromoelectric and magnetic fields need to be discretized. For the chromoelectric field we use \cite{Caron-Huot:2009ncn}
\begin{linenomath}\begin{align}
    E_i(\tau,\mathbf{x})=U_i(\tau,\mathbf{x})U_4(\tau,\mathbf{x}+a\hat{\mathbf{e}_i})-U_4(\tau,\mathbf{x})U_i(\tau,\mathbf{x}+a\hat{\mathbf{e}}_4),
\end{align}\end{linenomath}
where $U_\mu(\tau,\mathbf{x})$ is the link variable with lattice spacing $a$.
Similarly, we use the following discretization for $B$-field
\begin{linenomath}\begin{align}
    B_i(\tau,\mathbf{x})=\epsilon _{ijk}U_j(\tau,\mathbf{x})U_k(\tau,\mathbf{x}+a\hat{\mathbf{e}}_j),
    \label{eq:B_field_loop}
\end{align}\end{linenomath}
where we now obtain a loop within the spatial plane which lies perpendicularly to the spatial direction $i$ of the $B_i$-field component. By expanding the link variables in a small lattice spacing in Eq. \eqref{eq:B_field_loop} it can be shown that the discretized $B$-field corresponds to the chromomagnetic components of the field strength tensor at leading order.

 The chromoelectric correlator calculated on the lattice needs to be renormalized
 \cite{Christensen:2016wdo}. For the chromomagnetic correlator, renormalization is
 needed even in the continuum theory \cite{Laine:2021uzs}.
Since gradient flow renormalizes gauge invariant observables automatically \cite{Luscher:2010iy}, no renormalization is needed for measurements at finite flow time.

\begin{table}
	\centering
     \begin{subtable}[h]{0.35\textwidth}
         \centering
          \begin{tabular}{c|c|c|c}
          $N_s$ & $N_t$ & $\beta$ & $N_{\mathrm{conf}}$ \\ 
          \hline \hline
          48 & 16 & 6.872 & 990 \\ 
          \hline 
          48 & 20 & 7.044 & 3660 \\ 
          \hline 
          48 & 24 & 7.192 & 4200 \\ 
          \hline 
          56 & 28 & 7.321 & 4080
          \end{tabular} 
          \caption{$T=1.5T_c$}
     \end{subtable}
     \hspace{1em}
     \begin{subtable}[h]{0.35\textwidth}
         \centering
          \begin{tabular}{c|c|c|c}
          $N_s$ & $N_t$ & $\beta$ & $N_{\mathrm{conf}}$ \\ 
          \hline \hline
          48 & 16 & 14.443 & 990 \\ 
          \hline 
          48 & 20 & 14.635 & 990 \\ 
          \hline 
          48 & 24 & 14.792 & 1500 \\ 
          \hline 
          56 & 28 & 14.925 & 1950 \\ 
          \end{tabular} 
          \caption{$T=10^4T_c$}
     \end{subtable}
     \caption{The lattice parameters used in our calculations for $1.5T_c$
     and $10^4 T_c$. Here $T_c$ is the critical temperature.}
     \label{tab:lattice_setup}
\end{table}
In this study we focus on two temperatures $1.5T_c$ and $10^4T_c$ on quenched lattices as listed in table \ref{tab:lattice_setup}.  Here the critical temperature $T_c$ 
and the temperature values are determined using the scale setting procedure of Ref. \cite{Francis:2015lha}. The ensembles are generated with  Wilson gauge action using the heatbath and overrelaxation algorithm. We compute the field correlators using the gradient flow algorithm \cite{Luscher:2010iy, Bazavov:2021pik} with Symanzik action. In order to obtain a dimensionless and tree level improved quantity we normalize the final measured correlator values with the leading order lattice correlator \cite{Altenkort:2020fgs} at zero flow time
\begin{linenomath}\begin{align}
    G^{\scriptscriptstyle\text{norm}}=\frac{G^{\substack{\scriptscriptstyle\text{LO}\\ \scriptscriptstyle\text{latt}}}(\tau, \tau_F=0)}{g^2C_F},
\end{align}\end{linenomath}
where $\tau_F$ is the finite flow time. We assume the leading order of the chromoelectric and magnetic correlators to be the same. We obtain a tree level improved quantity by using the perturbative lattice correlator as normalization instead of the perturbative continuum correlator.

The final renormalized values of the correlators are obtained in a zero flow time limit. However, the zero flow time limit can only be taken after the continuum 
extrapolation is performed. Therefore, we calculate the correlators at different lattice spacings and then perform the continuum extrapolation. Afterwards, we can perform the zero flow time limit. 
\begin{figure}
	\centering
     \begin{subfigure}[h]{0.45\textwidth}
         \centering
          \includegraphics[scale=0.5]{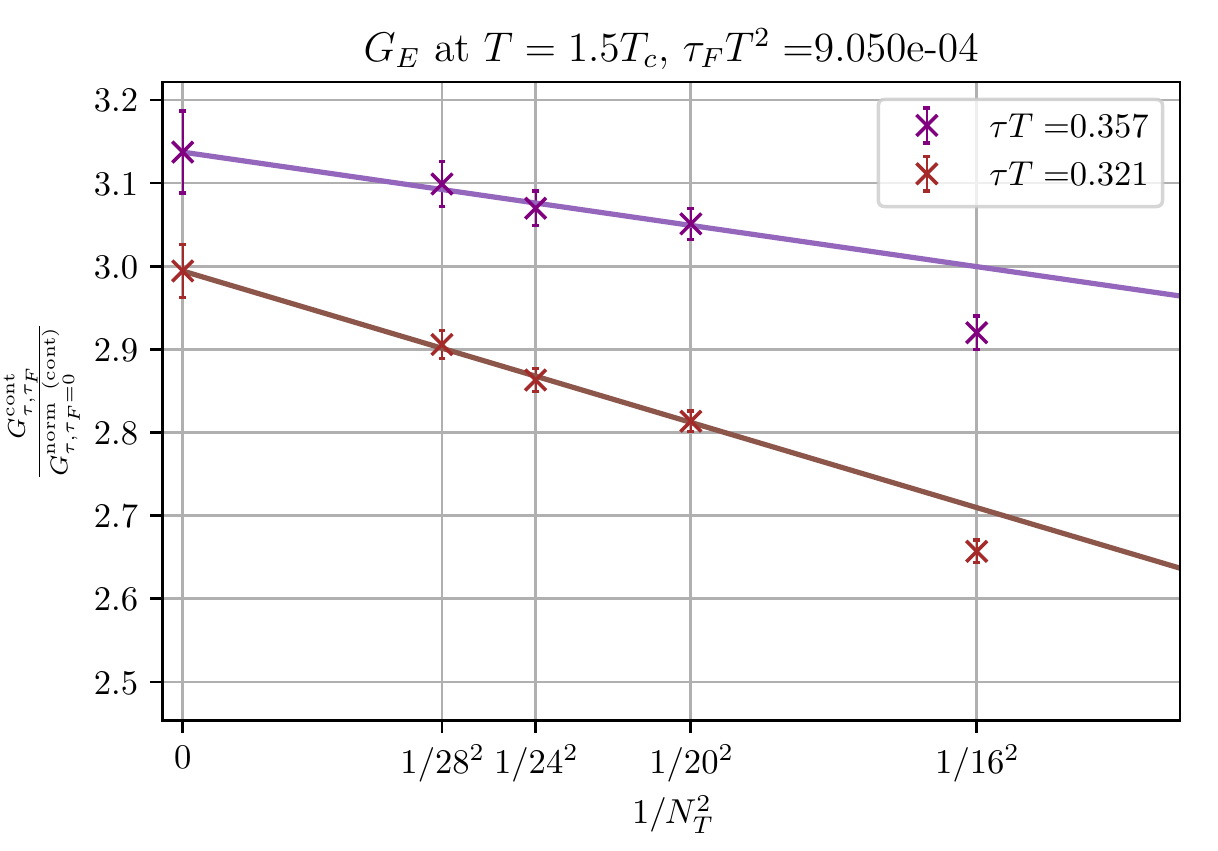}
     \end{subfigure}
     \begin{subfigure}[h]{0.45\textwidth}
         \centering
          \includegraphics[scale=0.5]{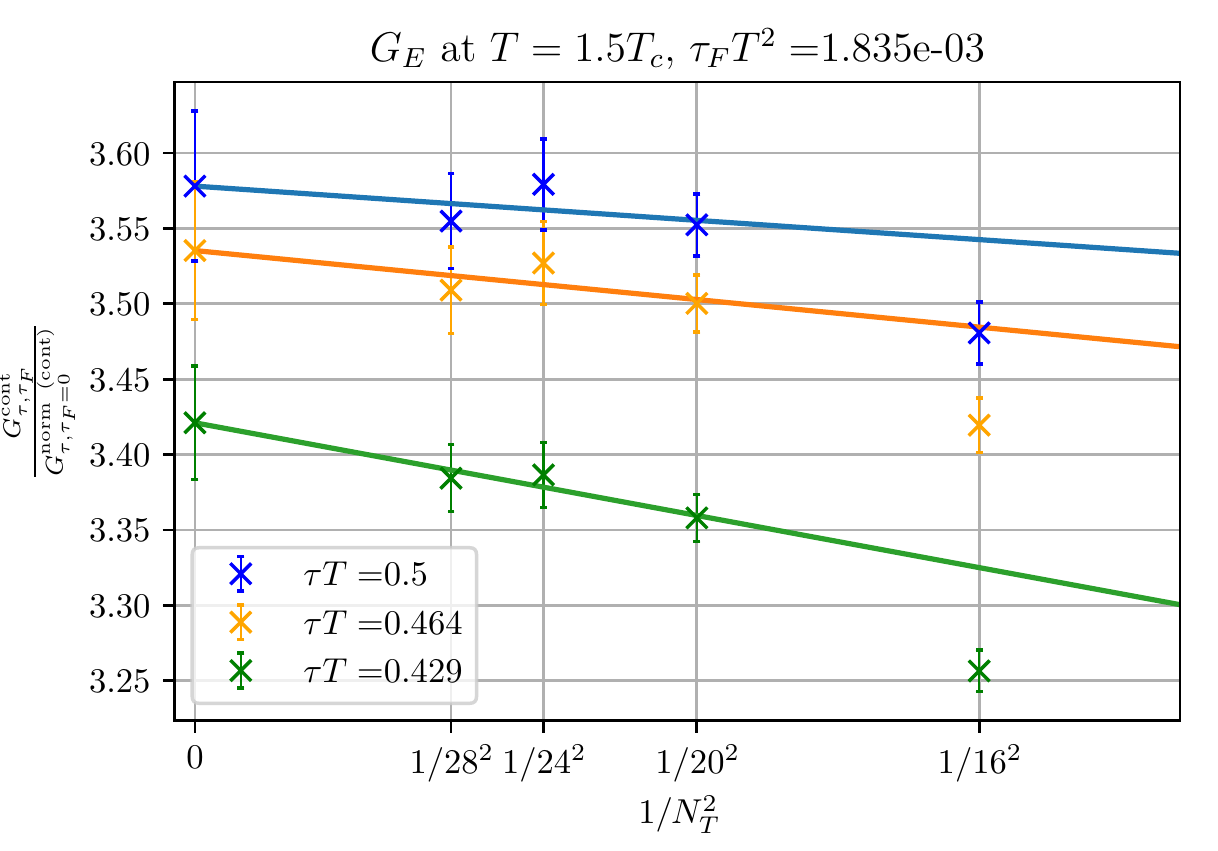}
     \end{subfigure}
     \vfill
     \begin{subfigure}[h]{0.45\textwidth}
         \centering
          \includegraphics[scale=0.5]{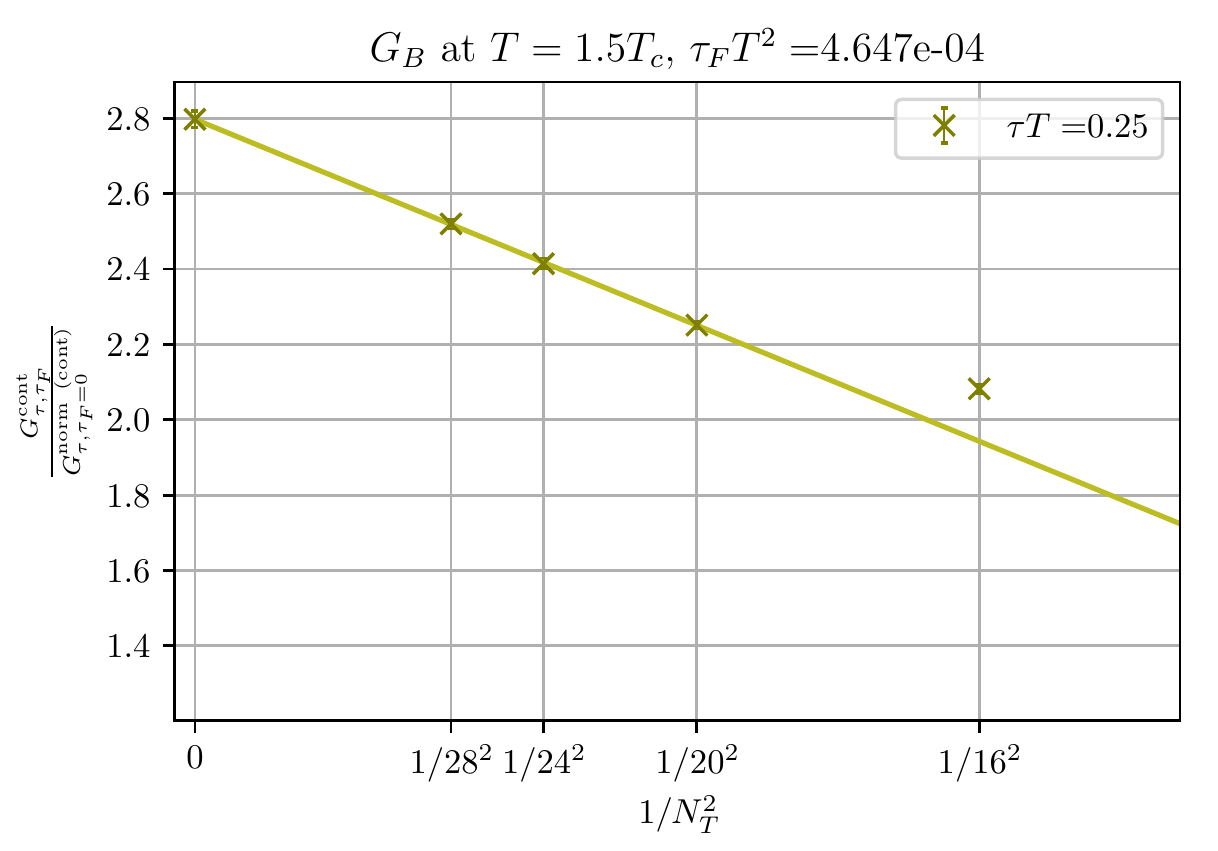}
     \end{subfigure}
     \begin{subfigure}[h]{0.45\textwidth}
         \centering
          \includegraphics[scale=0.5]{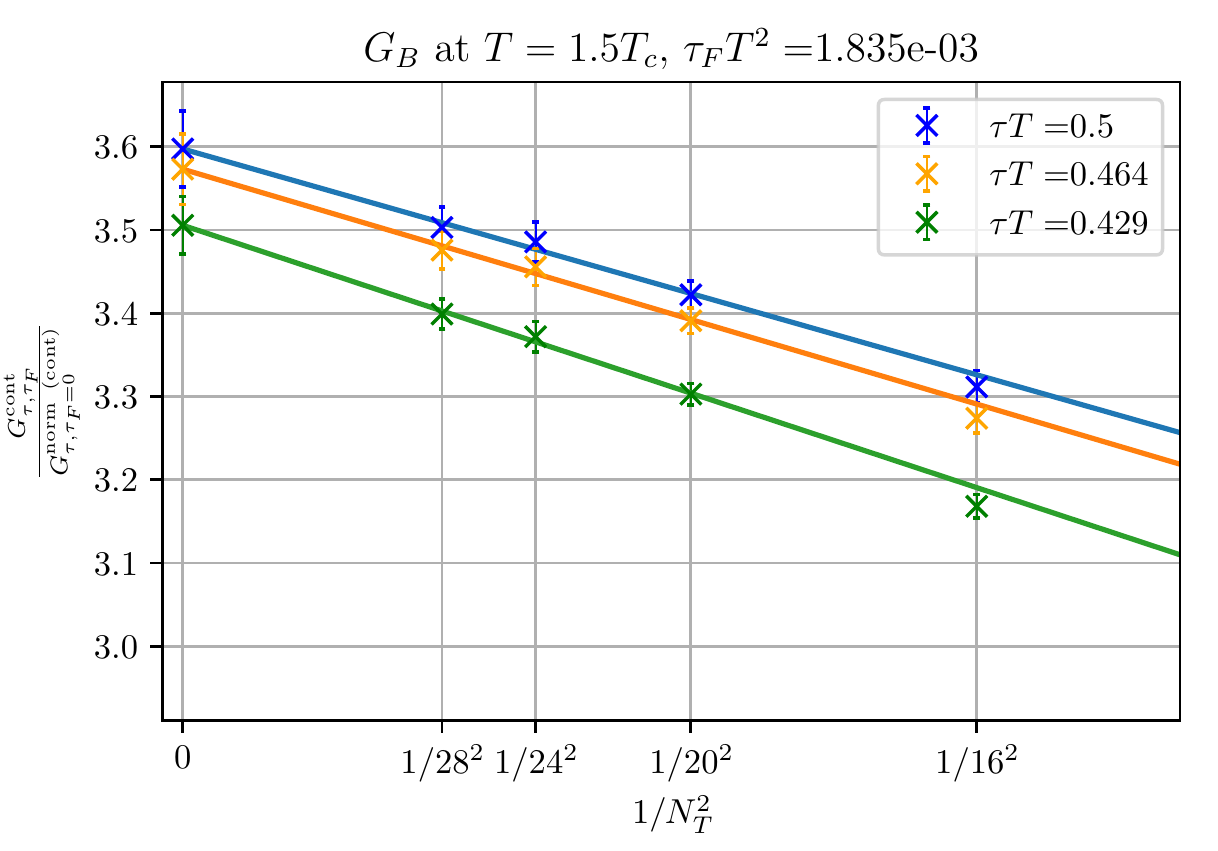}
     \end{subfigure}
     \caption{Examples for the continuum limits of the chromoelectric and chromomagnetic correlators at four different flow times at $T=1.5T_c$.}
     \label{fig:Ge_cont_limit_cold}
\end{figure}
\begin{figure}
	\centering
     \begin{subfigure}[h]{0.45\textwidth}
         \centering
          \includegraphics[scale=0.5]{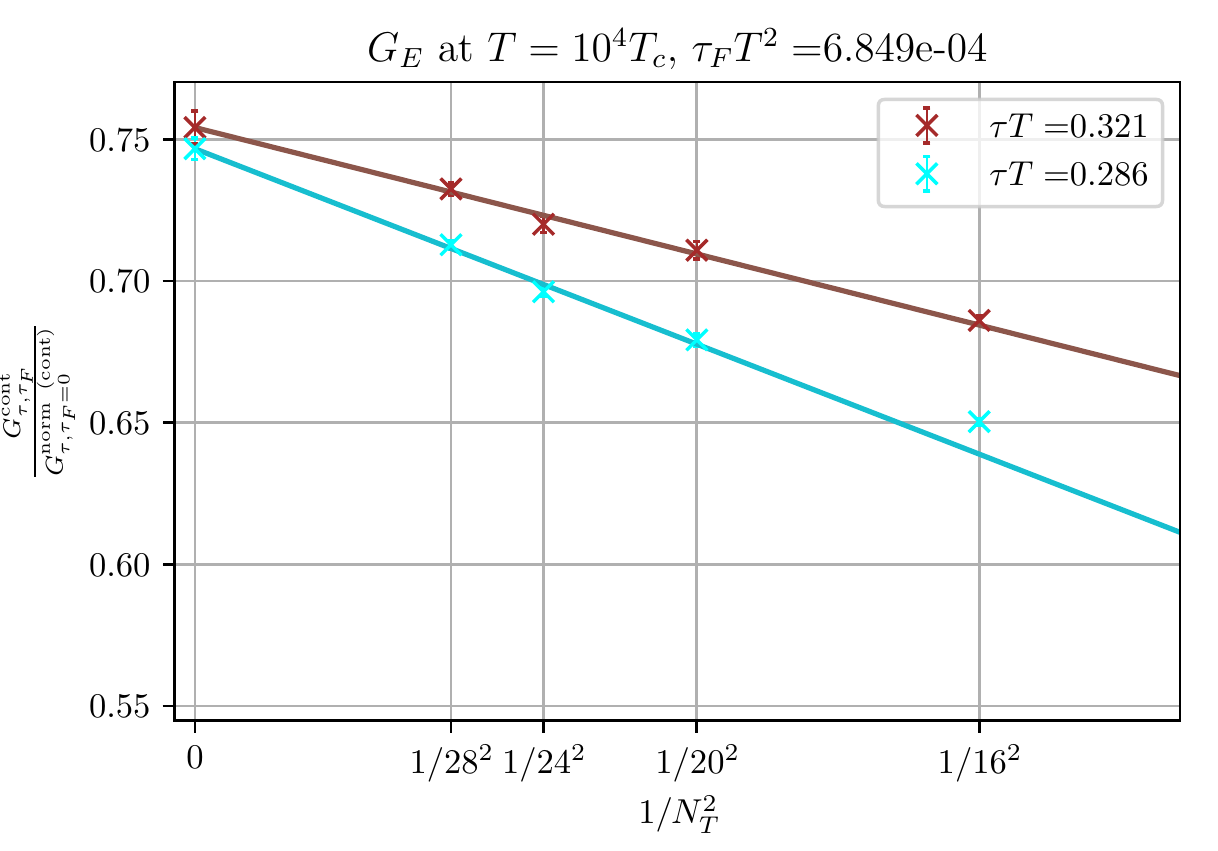}
     \end{subfigure}
     \begin{subfigure}[h]{0.45\textwidth}
         \centering
          \includegraphics[scale=0.5]{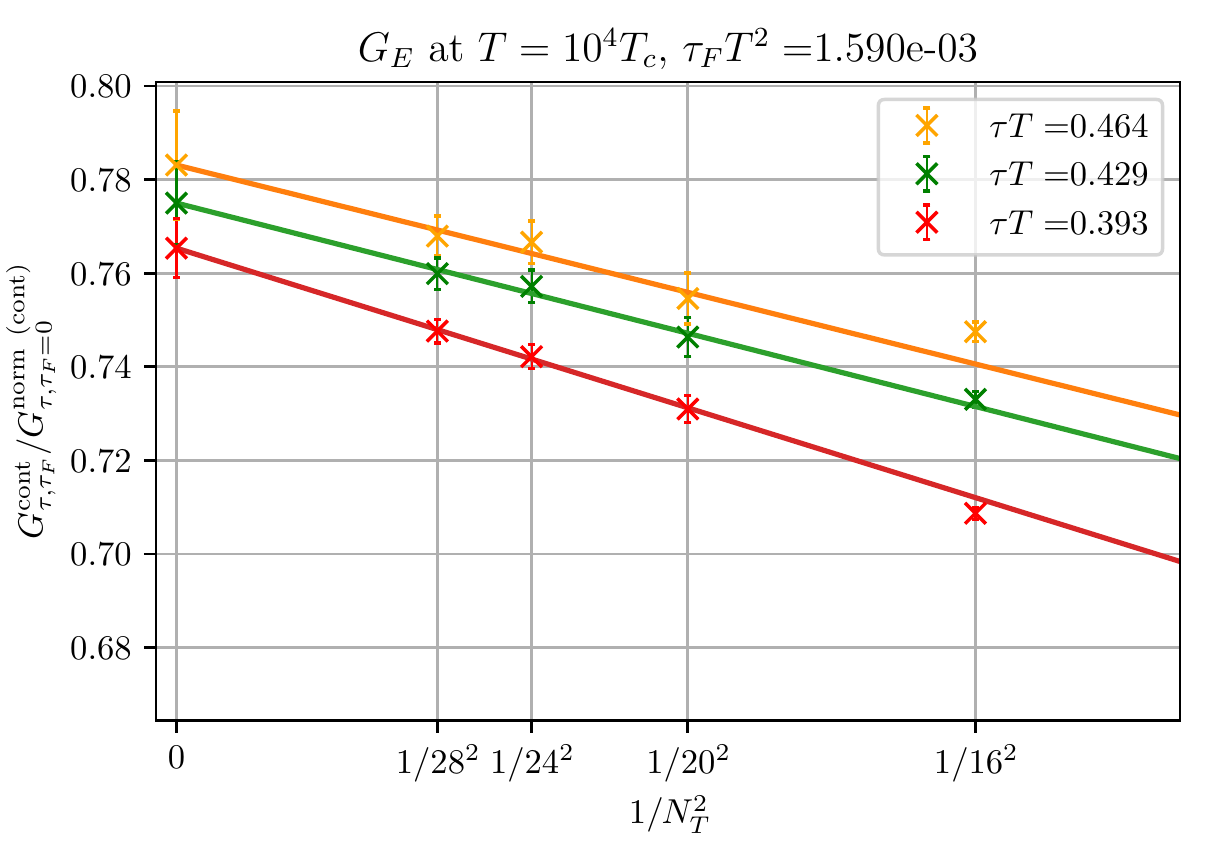}
     \end{subfigure}
     \vfill
     \begin{subfigure}[h]{0.45\textwidth}
         \centering
          \includegraphics[scale=0.5]{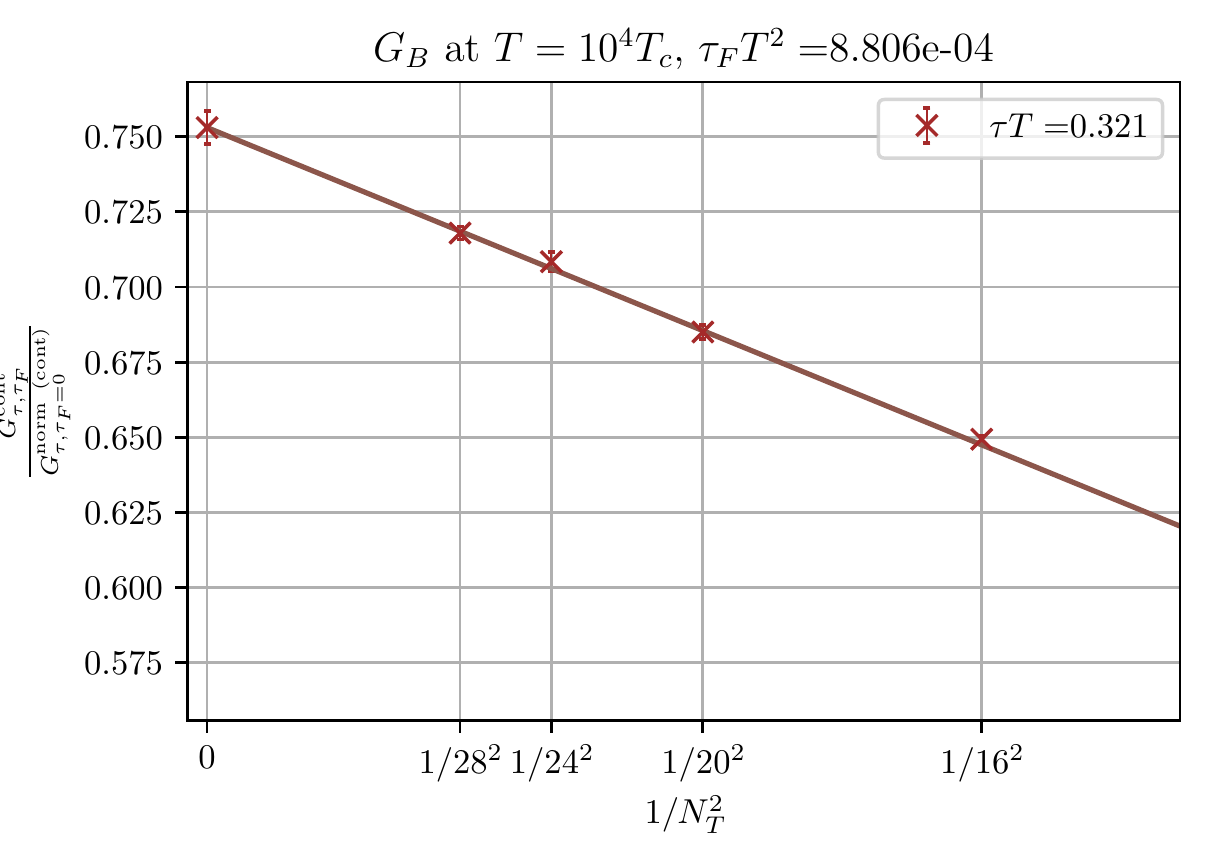}
     \end{subfigure}
     \begin{subfigure}[h]{0.45\textwidth}
         \centering
          \includegraphics[scale=0.5]{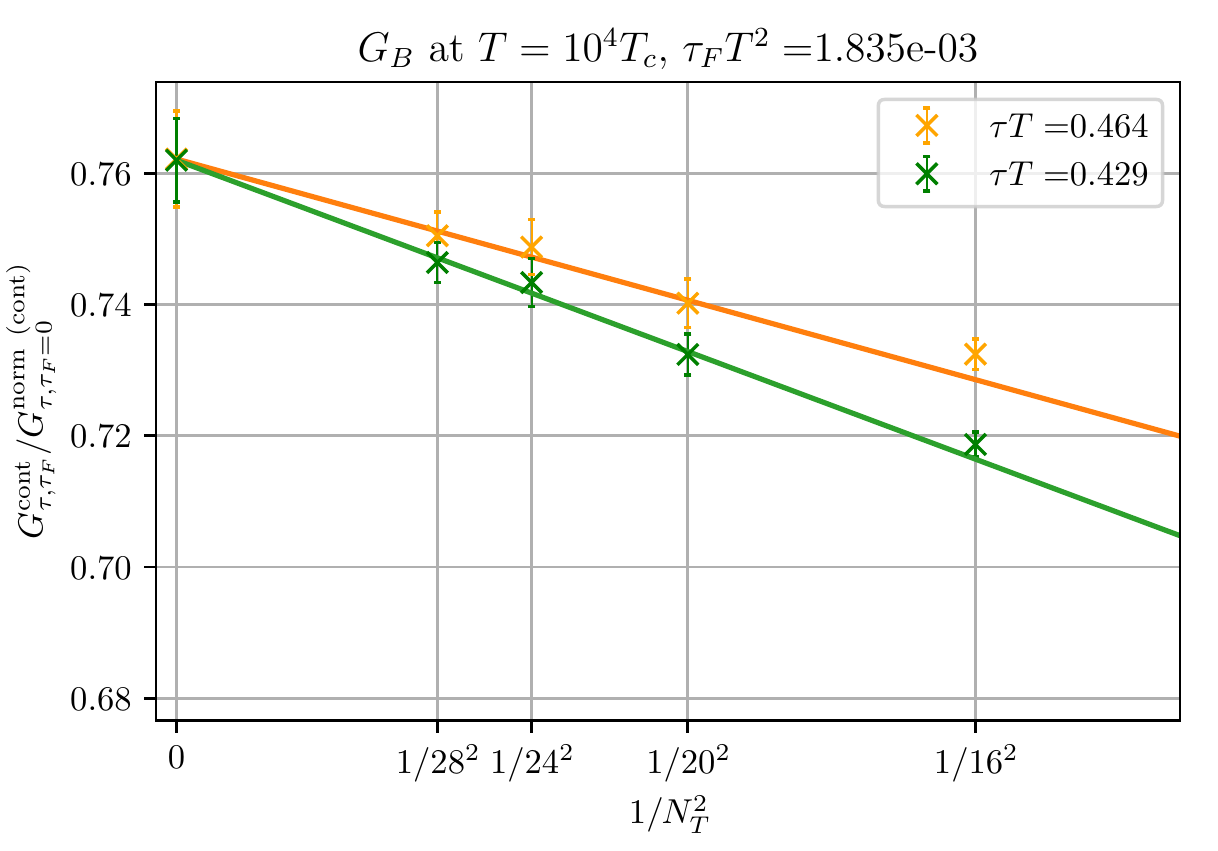}
     \end{subfigure}
     \caption{Examples for the continuum limits of the chromoelectric and chromomagnetic correlators at four different flow times at $T=10^4T_c$.}
     \label{fig:Ge_cont_limit_hot}
\end{figure}
To obtain the continuum limit, we interpolate the data in $\tau T$ with cubic splines, using natural boundary conditions at $\tau T=0$ (second derivative equal to zero) and symmetric boundary conditions at $\tau T=0.5$ (first derivative equal to zero) in order to have data of the same $\tau T$ axis on all lattices, and perform linear extrapolations over $1/N_t^2$ to the continuum ($N_t\rightarrow\infty$) using lattices with temporal extents of \mbox{$N_t=20, 24, 28$}. Figs. \ref{fig:Ge_cont_limit_cold} and \ref{fig:Ge_cont_limit_hot} show some typical
examples of continuum extrapolations for the chromoelectric and magnetic fields respectively. In these figures, we also show the $N_t=16$ data points.
It appears that the $N_t=16$ data are not in $1/N_t^2$ scaling regime. Therefore, we do not include these
in the continuum extrapolations.
We expect the spatial size dependence to be negligible \cite{Brambilla:2020siz}.

\begin{figure}
	\centering
     \begin{subfigure}[h]{0.45\textwidth}
         \centering
          \includegraphics[scale=0.45]{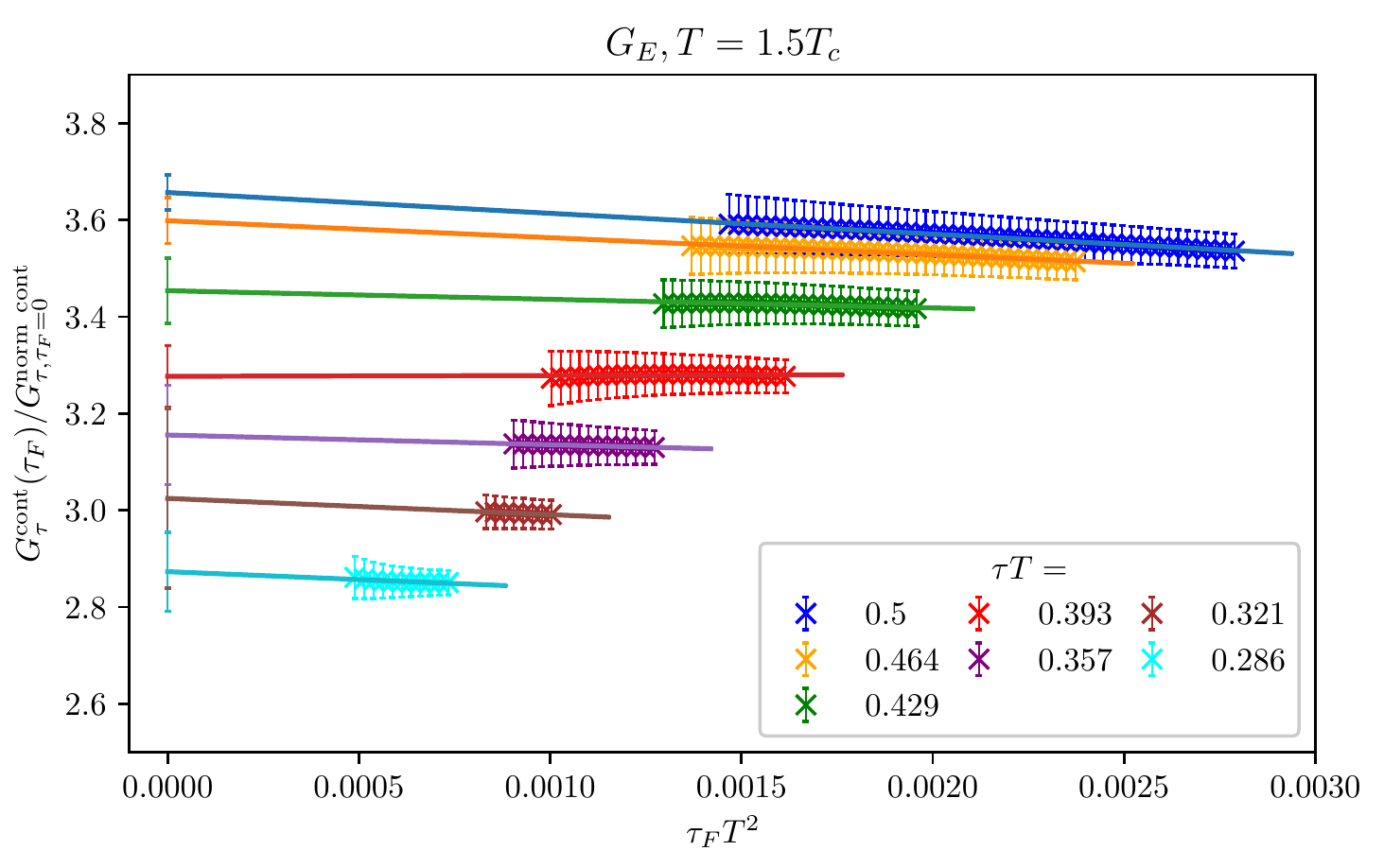}
     \end{subfigure}
     \begin{subfigure}[h]{0.45\textwidth}
         \centering
          \includegraphics[scale=0.45]{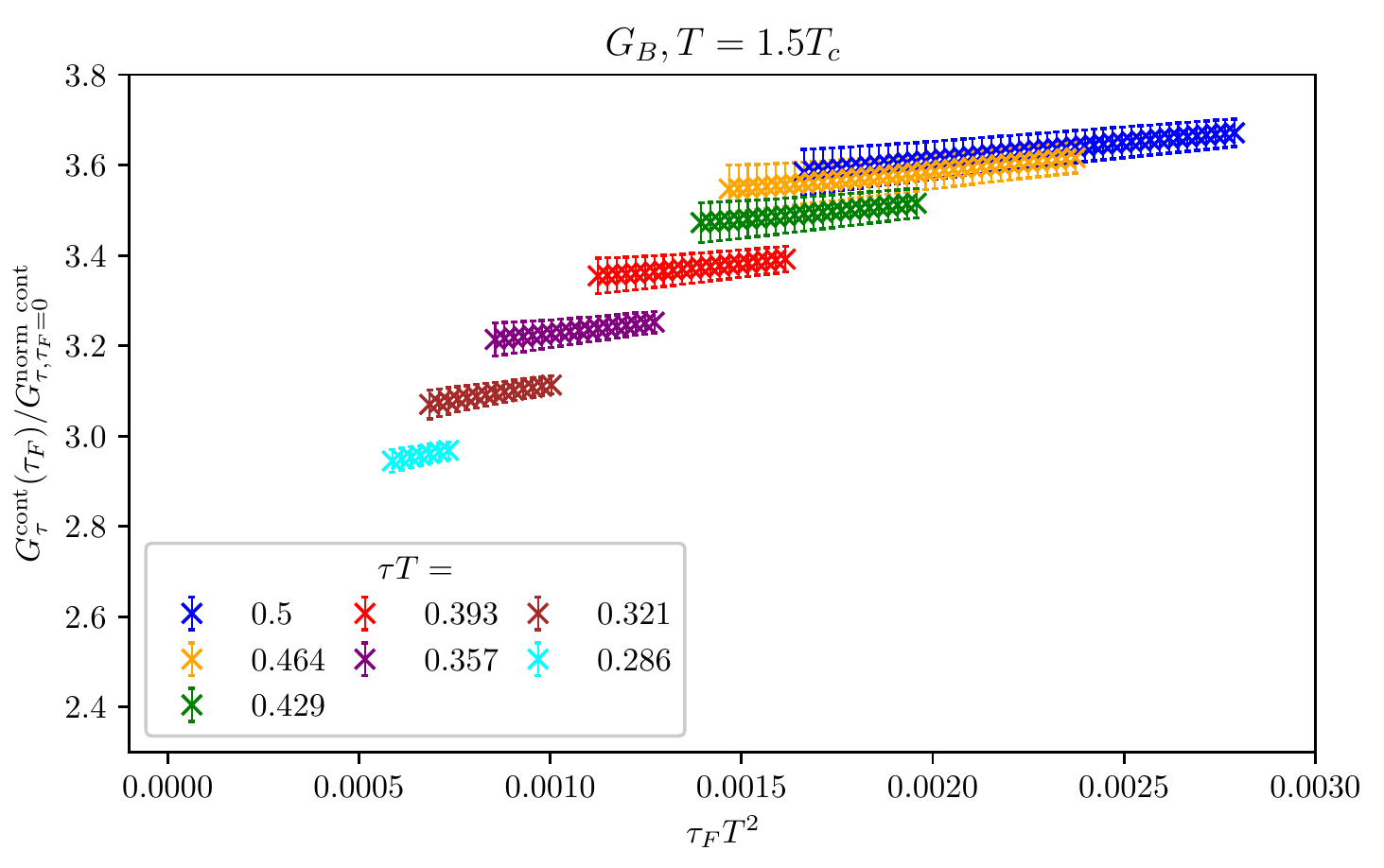}
     \end{subfigure}
     \caption{The final results of the continuum limit for the chromoelectric and magnetic correlators at $T=1.5T_c$.}
     \label{fig:zero_flow time_limit_cold}
\end{figure}
In Fig. \ref{fig:zero_flow time_limit_cold} we
show the continuum extrapolated
results for $T=1.5T_c$. For the chromoelectric correlator we perform a linear extrapolation
in $\tau_F T^2$, which is also shown in Fig. \ref{fig:zero_flow time_limit_cold}.
When performing the extrapolation we considered flow time $\tau_F$ in a restricted range 
\begin{linenomath}\begin{align}
    a\leq\sqrt{8\tau_F}\leq\frac{\tau-a}{3}.
    \label{eq:flow time_limits}
\end{align}\end{linenomath}
Here $\sqrt{8\tau_F}$ is often called the flow radius. 
The above constrain for the flow time comes from the fact that for this range 
the perturbative flowed correlator deviates less than \SI{1}{\percent} 
from the unflowed correlator \cite{Altenkort:2020fgs}.
For larger flow times the dependence
on $\tau_F T^2$ is nonlinear and the corresponding data cannot be used for the 
$\tau_F \rightarrow 0$ extrapolations.
It is obvious from Fig. \ref{fig:zero_flow time_limit_cold} that the flow time dependence
of the chromoelectric and chromomagnetic correlators is quite different. It remains to bee
seen if this difference in the flow time dependence is related to the anomalous dimension
of the chromomagnetic correlator \cite{Laine:2021uzs}. 
\begin{figure}
	\centering
     \begin{subfigure}[h]{0.45\textwidth}
         \centering
          \includegraphics[scale=0.45]{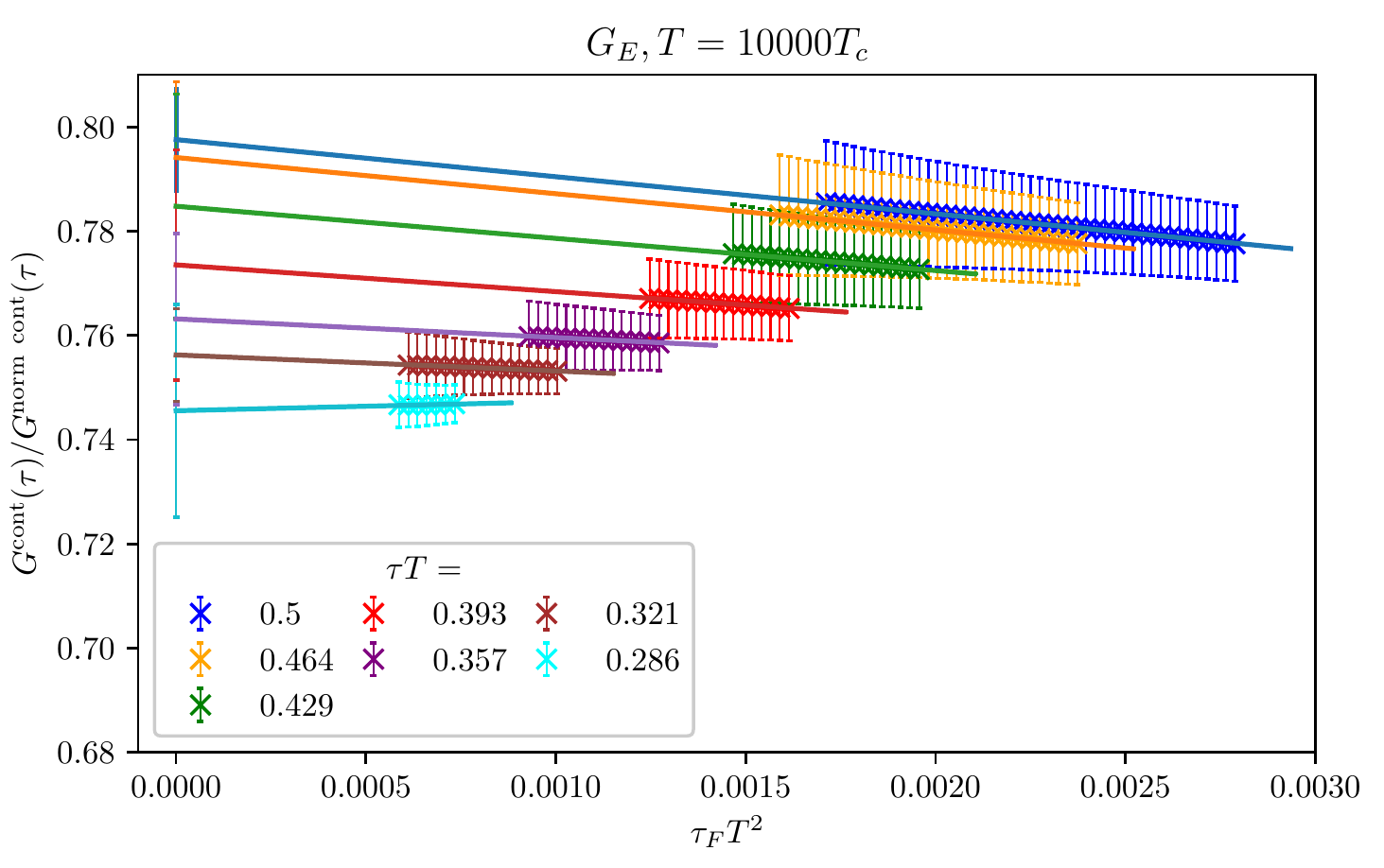}
     \end{subfigure}
     \begin{subfigure}[h]{0.45\textwidth}
         \centering
          \includegraphics[scale=0.45]{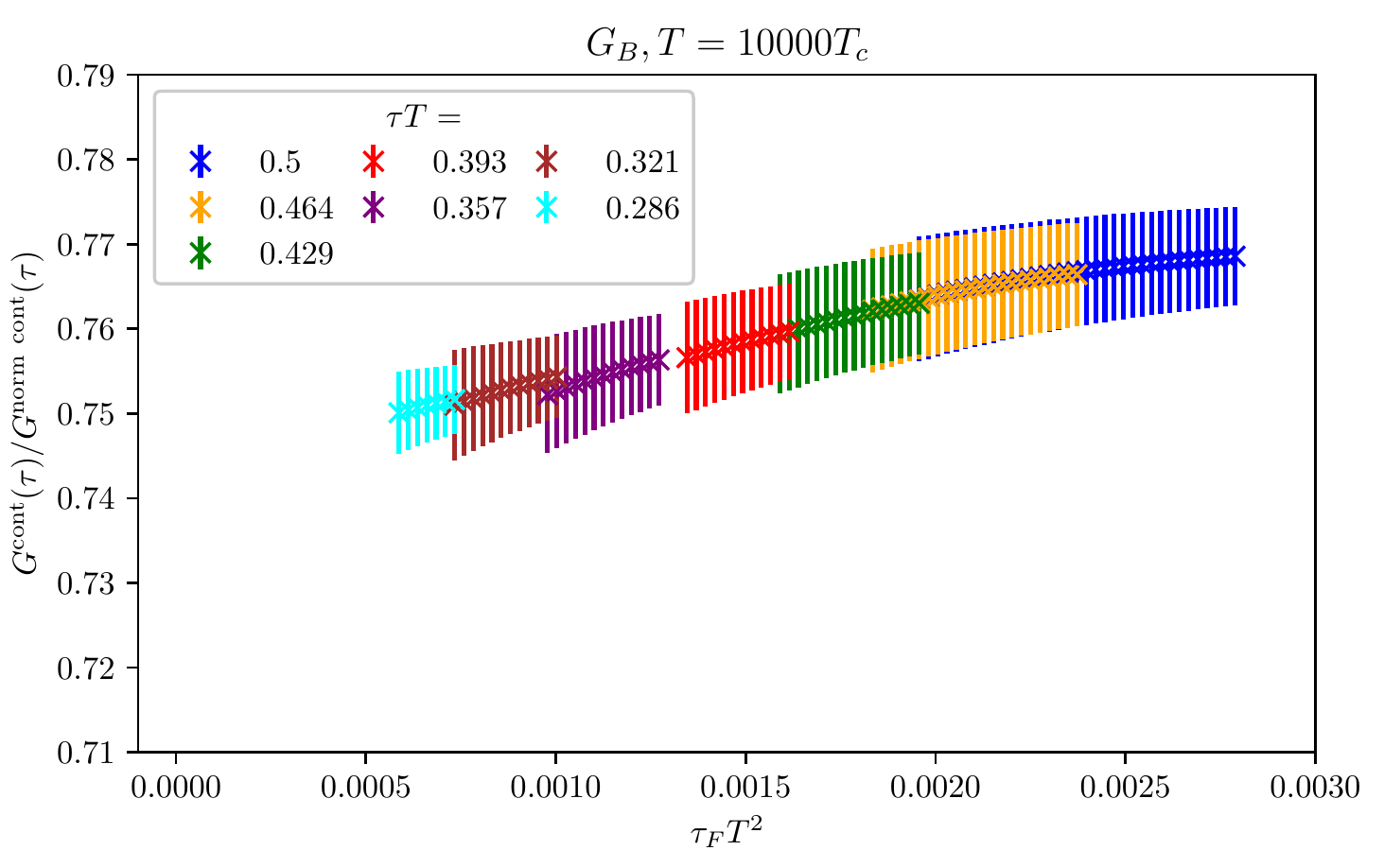}
     \end{subfigure}
     \caption{The final results of the continuum limit for the chromoelectric and magnetic correlators at $T=10^4T_c$.}
     \label{fig:zero_flow time_limit_hot}
\end{figure}
Fig. \ref{fig:zero_flow time_limit_hot} shows the final results of the continuum limits at $T=10^4T_c$. We perform the same zero flow time procedure as for the $T=1.5T_c$ case as indicated in Fig. \ref{fig:zero_flow time_limit_hot}. We obtain the same flow time behaviour of the correlators for $\tau_F\rightarrow 0$. 

\begin{figure}
	\centering
     \begin{subfigure}[h]{0.45\textwidth}
         \centering
          \includegraphics[scale=0.45]{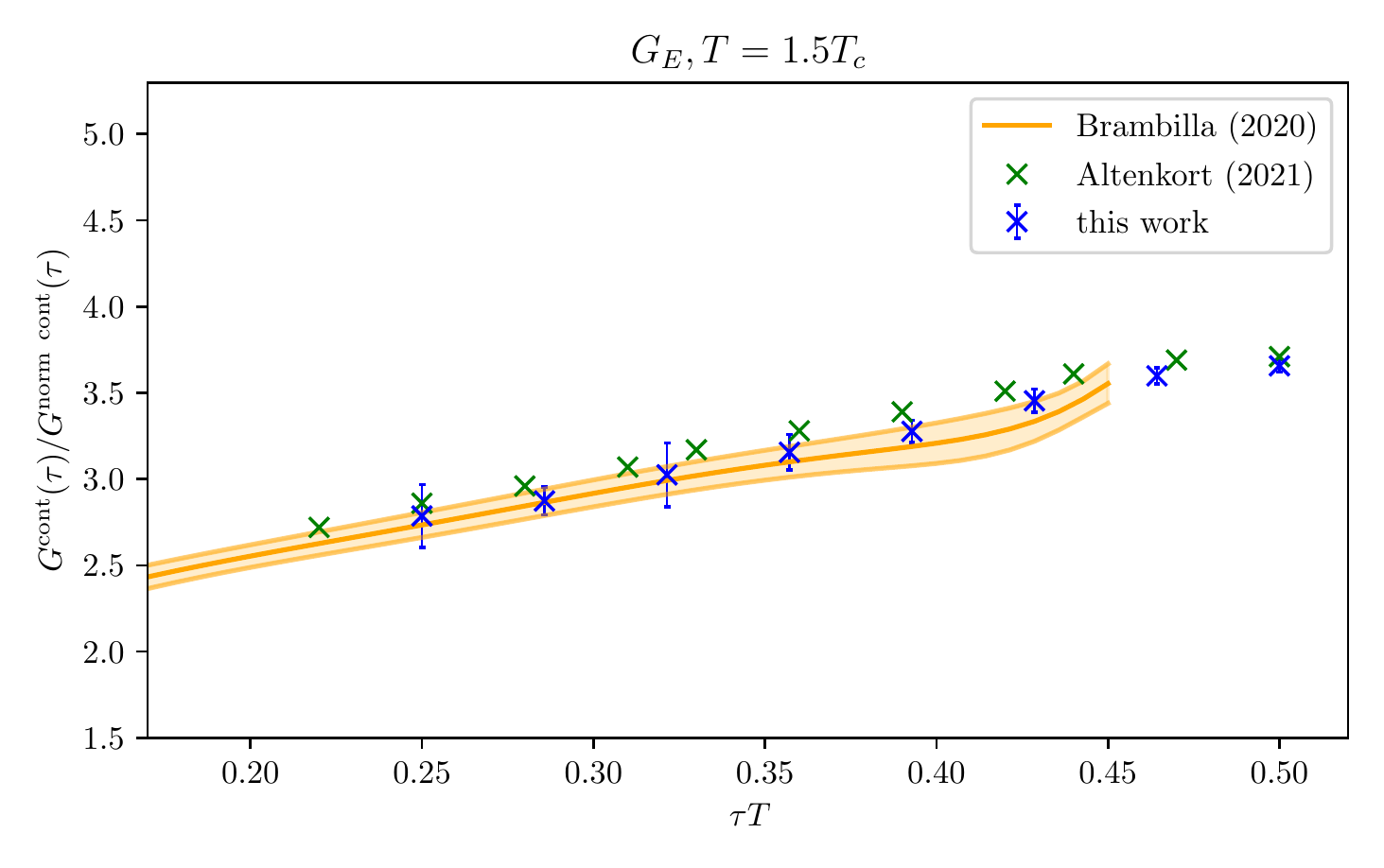}
     \end{subfigure}
     \begin{subfigure}[h]{0.45\textwidth}
         \centering
          \includegraphics[scale=0.45]{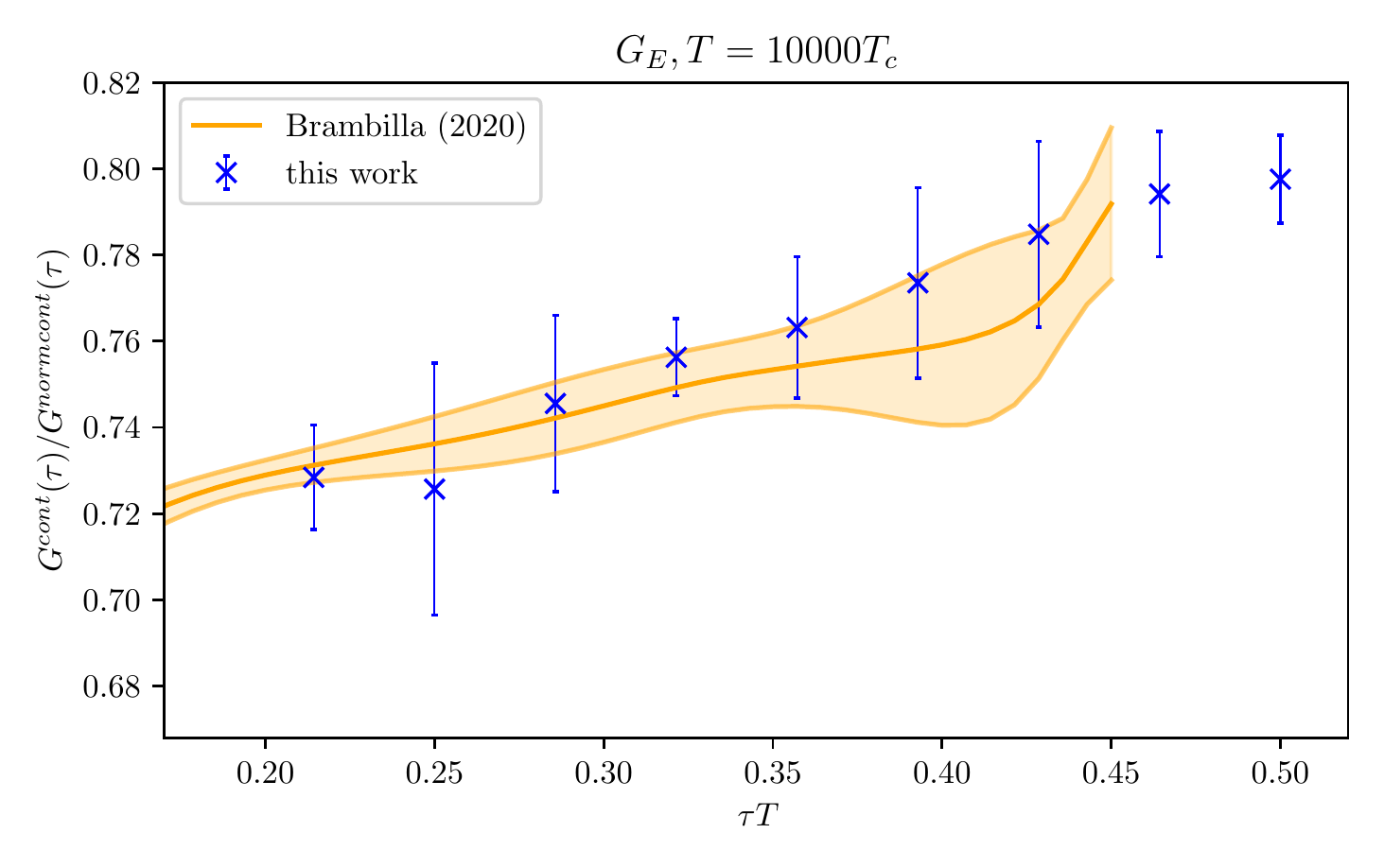}
     \end{subfigure}
     \caption{Final results on $G_E$ in the limit of zero flow time compared with previous results \cite{Altenkort:2020fgs} based on gradient flow
     and on the calculations with multilevel algorithm \cite{Brambilla:2020siz}.}
     \label{fig:final_results_Ge}
\end{figure}

In Fig. \ref{fig:final_results_Ge} we show the final results 
for the chromoelectric correlator as function of $\tau$ in the 
zero flow time limit.   Our results are compared with results from Ref. \cite{Altenkort:2020fgs} at $T=1.5T_c$,
which also rely on gradient flow, 
and with results from Ref. \cite{Brambilla:2020siz}, which are calculated with a multilevel approach at $T=1.5T_c$ and $T=10^4T_c$. The multilevel results are renormalized by the 1-loop renormalization constant from Ref. \cite{Christensen:2016wdo}. We see that our results agree with the previous calculations at both temperatures, hence, we can conclude that the gradient flow approach serves as a non-perturbative renormalization method.

\begin{figure}
	\centering
     \begin{subfigure}[h]{0.45\textwidth}
         \centering
          \includegraphics[scale=0.45]{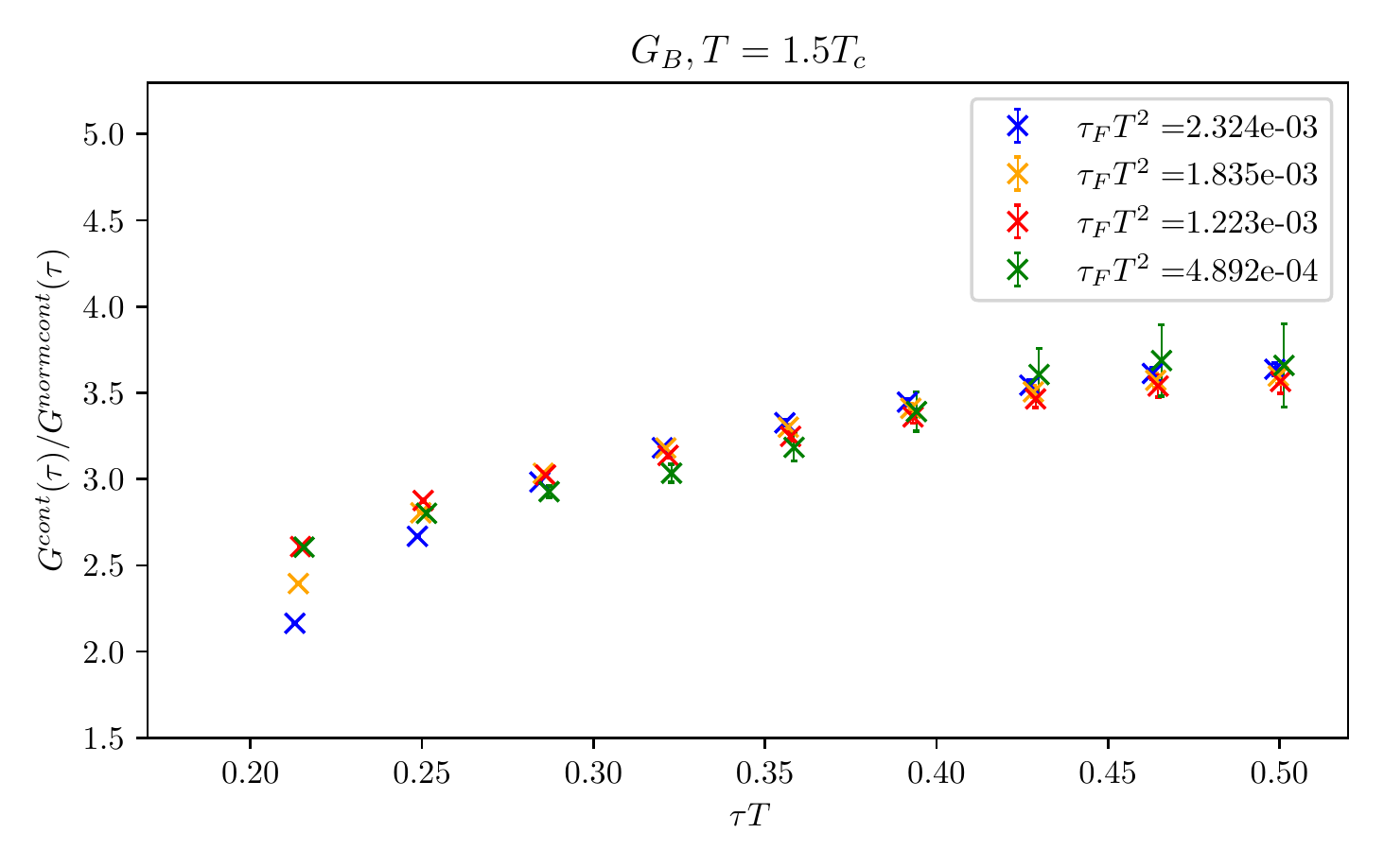}
     \end{subfigure}
     \begin{subfigure}[h]{0.45\textwidth}
         \centering
          \includegraphics[scale=0.45]{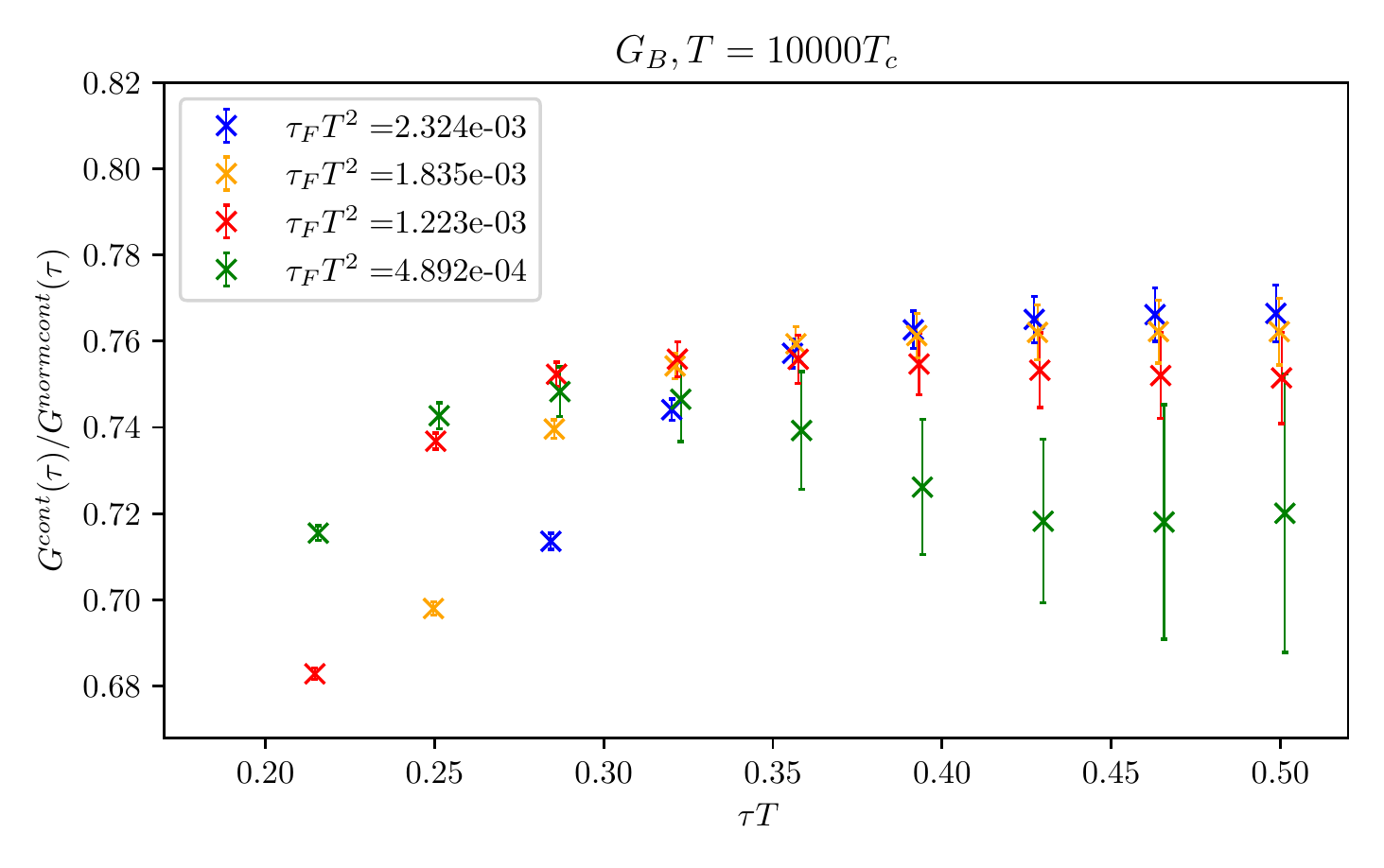}
     \end{subfigure}
     \caption{The chromomagnetic correlator as function of $\tau$
     obtained for different flow times, $\tau_F$.}
     \label{fig:final_results_Gb}
\end{figure}
For the chromomagnetic correlator the zero flow time limit cannot be taken
because of the anomalous dimension \cite{Laine:2021uzs}. Therefore,
in Fig. \ref{fig:final_results_Gb} we show the chromomagnetic correlator for
four different flow times. We see that $G_B$ is almost $\tau_F$-independent for $\tau T\geq 0.25$ at $T=1.5T_c$.
For $T=10^4T_c$ the chromomagnetic correlator is roughly
flow time independent for $\tau T\geq 0.3$.

\section{Conclusions}

In this contribution we studied the chromoelectric and chromomagnetic correlators in quenched QCD at two temperatures, $1.5T_c$ and $10^4T_c$.
These correlators are interesting as they encode information on
the heavy quark diffusion coefficient. We used the gradient flow to
reduce the noise in the lattice calculations of these correlators as well
as to obtain the renormalized results. We have found that the flow time
dependence of the chromoelectric and chromomagnetic correlators is quite
different for small flow times. For the chromoelectric correlator we 
performed the zero flow time extrapolation and found that for both
temperatures the zero flow time extrapolated results agree with the previously published ones.

\acknowledgments{The simulations were performed using the MILC code. The simulations have been carried out on the computing facilities of the Computational Center for Particle and Astrophysics (C2PAP) of the cluster of excellence ORIGINS that is funded by the Deutsche Forschungsgemeinschaft under Germany's Excellence Strategy EXC-2094-390783311.
PP was supported by the U.S. Department of Energy through Contract No. DE-SC0012704.}
\bibliographystyle{JHEP}
\bibliography{bibliography.bib}

\providecommand{\href}[2]{#2}\begingroup\raggedright\begin{thebibliography}{10}

\bibitem{Pasechnik:2016wkt}
R.~Pasechnik and M.~\v{S}umbera, \emph{{Phenomenological Review on
  Quark\textendash{}Gluon Plasma: Concepts vs. Observations}},
  \href{https://doi.org/10.3390/universe3010007}{\emph{Universe} {\bfseries 3}
  (2017) 7} [\href{https://arxiv.org/abs/1611.01533}{{\ttfamily 1611.01533}}].

\bibitem{Moore:2004tg}
G.D.~Moore and D.~Teaney, \emph{{How much do heavy quarks thermalize in a heavy
  ion collision?}},
  \href{https://doi.org/10.1103/PhysRevC.71.064904}{\emph{Phys. Rev. C}
  {\bfseries 71} (2005) 064904}
  [\href{https://arxiv.org/abs/hep-ph/0412346}{{\ttfamily hep-ph/0412346}}].

\bibitem{Svetitsky:1987gq}
B.~Svetitsky, \emph{{Diffusion of charmed quarks in the quark-gluon plasma}},
  \href{https://doi.org/10.1103/PhysRevD.37.2484}{\emph{Phys. Rev. D}
  {\bfseries 37} (1988) 2484}.

\bibitem{Caron-Huot:2008dyw}
S.~Caron-Huot and G.D.~Moore, \emph{{Heavy quark diffusion in QCD and N=4 SYM
  at next-to-leading order}},
  \href{https://doi.org/10.1088/1126-6708/2008/02/081}{\emph{JHEP} {\bfseries
  02} (2008) 081} [\href{https://arxiv.org/abs/0801.2173}{{\ttfamily
  0801.2173}}].

\bibitem{Casalderrey-Solana:2006fio}
J.~Casalderrey-Solana and D.~Teaney, \emph{{Heavy quark diffusion in strongly
  coupled N=4 Yang-Mills}},
  \href{https://doi.org/10.1103/PhysRevD.74.085012}{\emph{Phys. Rev. D}
  {\bfseries 74} (2006) 085012}
  [\href{https://arxiv.org/abs/hep-ph/0605199}{{\ttfamily hep-ph/0605199}}].

\bibitem{Brambilla:2017zei}
N.~Brambilla, M.A.~Escobedo, J.~Soto and A.~Vairo, \emph{{Heavy quarkonium
  suppression in a fireball}},
  \href{https://doi.org/10.1103/PhysRevD.97.074009}{\emph{Phys. Rev. D}
  {\bfseries 97} (2018) 074009}
  [\href{https://arxiv.org/abs/1711.04515}{{\ttfamily 1711.04515}}].

\bibitem{Meyer:2010tt}
H.B.~Meyer, \emph{{The errant life of a heavy quark in the quark-gluon
  plasma}}, \href{https://doi.org/10.1088/1367-2630/13/3/035008}{\emph{New J.
  Phys.} {\bfseries 13} (2011) 035008}
  [\href{https://arxiv.org/abs/1012.0234}{{\ttfamily 1012.0234}}].

\bibitem{Francis:2011gc}
A.~Francis, O.~Kaczmarek, M.~Laine and J.~Langelage, \emph{{Towards a
  non-perturbative measurement of the heavy quark momentum diffusion
  coefficient}}, \href{https://doi.org/10.22323/1.139.0202}{\emph{PoS}
  {\bfseries LATTICE2011} (2011) 202}
  [\href{https://arxiv.org/abs/1109.3941}{{\ttfamily 1109.3941}}].

\bibitem{Banerjee:2011ra}
D.~Banerjee, S.~Datta, R.~Gavai and P.~Majumdar, \emph{{Heavy Quark Momentum
  Diffusion Coefficient from Lattice QCD}},
  \href{https://doi.org/10.1103/PhysRevD.85.014510}{\emph{Phys. Rev. D}
  {\bfseries 85} (2012) 014510}
  [\href{https://arxiv.org/abs/1109.5738}{{\ttfamily 1109.5738}}].

\bibitem{Francis:2015daa}
A.~Francis, O.~Kaczmarek, M.~Laine, T.~Neuhaus and H.~Ohno,
  \emph{{Nonperturbative estimate of the heavy quark momentum diffusion
  coefficient}}, \href{https://doi.org/10.1103/PhysRevD.92.116003}{\emph{Phys.
  Rev. D} {\bfseries 92} (2015) 116003}
  [\href{https://arxiv.org/abs/1508.04543}{{\ttfamily 1508.04543}}].

\bibitem{Brambilla:2020siz}
N.~Brambilla, V.~Leino, P.~Petreczky and A.~Vairo, \emph{{Lattice QCD
  constraints on the heavy quark diffusion coefficient}},
  \href{https://doi.org/10.1103/PhysRevD.102.074503}{\emph{Phys. Rev. D}
  {\bfseries 102} (2020) 074503}
  [\href{https://arxiv.org/abs/2007.10078}{{\ttfamily 2007.10078}}].

\bibitem{Altenkort:2020fgs}
L.~Altenkort, A.M.~Eller, O.~Kaczmarek, L.~Mazur, G.D.~Moore and H.-T.~Shu,
  \emph{{Heavy quark momentum diffusion from the lattice using gradient flow}},
  \href{https://doi.org/10.1103/PhysRevD.103.014511}{\emph{Phys. Rev. D}
  {\bfseries 103} (2021) 014511}
  [\href{https://arxiv.org/abs/2009.13553}{{\ttfamily 2009.13553}}].

\bibitem{Bouttefeux:2020ycy}
A.~Bouttefeux and M.~Laine, \emph{{Mass-suppressed effects in heavy quark
  diffusion}}, \href{https://doi.org/10.1007/JHEP12(2020)150}{\emph{JHEP}
  {\bfseries 12} (2020) 150}
  [\href{https://arxiv.org/abs/2010.07316}{{\ttfamily 2010.07316}}].

\bibitem{Burnier:2010rp}
Y.~Burnier, M.~Laine, J.~Langelage and L.~Mether, \emph{{Colour-electric
  spectral function at next-to-leading order}},
  \href{https://doi.org/10.1007/JHEP08(2010)094}{\emph{JHEP} {\bfseries 08}
  (2010) 094} [\href{https://arxiv.org/abs/1006.0867}{{\ttfamily 1006.0867}}].

\bibitem{Caron-Huot:2009ncn}
S.~Caron-Huot, M.~Laine and G.D.~Moore, \emph{{A Way to estimate the heavy
  quark thermalization rate from the lattice}},
  \href{https://doi.org/10.1088/1126-6708/2009/04/053}{\emph{JHEP} {\bfseries
  04} (2009) 053} [\href{https://arxiv.org/abs/0901.1195}{{\ttfamily
  0901.1195}}].

\bibitem{Christensen:2016wdo}
C.~Christensen and M.~Laine, \emph{{Perturbative renormalization of the
  electric field correlator}},
  \href{https://doi.org/10.1016/j.physletb.2016.02.020}{\emph{Phys. Lett. B}
  {\bfseries 755} (2016) 316}
  [\href{https://arxiv.org/abs/1601.01573}{{\ttfamily 1601.01573}}].

\bibitem{Laine:2021uzs}
M.~Laine, \emph{{1-loop matching of a thermal Lorentz force}},
  \href{https://doi.org/10.1007/JHEP06(2021)139}{\emph{JHEP} {\bfseries 06}
  (2021) 139} [\href{https://arxiv.org/abs/2103.14270}{{\ttfamily
  2103.14270}}].

\bibitem{Luscher:2010iy}
M.~L\"uscher, \emph{{Properties and uses of the Wilson flow in lattice QCD}},
  \href{https://doi.org/10.1007/JHEP08(2010)071}{\emph{JHEP} {\bfseries 08}
  (2010) 071} [\href{https://arxiv.org/abs/1006.4518}{{\ttfamily 1006.4518}}].

\bibitem{Francis:2015lha}
A.~Francis, O.~Kaczmarek, M.~Laine, T.~Neuhaus and H.~Ohno, \emph{{Critical
  point and scale setting in SU(3) plasma: An update}},
  \href{https://doi.org/10.1103/PhysRevD.91.096002}{\emph{Phys. Rev. D}
  {\bfseries 91} (2015) 096002}
  [\href{https://arxiv.org/abs/1503.05652}{{\ttfamily 1503.05652}}].

\bibitem{Bazavov:2021pik}
A.~Bazavov and T.~Chuna, \emph{{Efficient integration of gradient flow in
  lattice gauge theory and properties of low-storage commutator-free Lie group
  methods}},  \href{https://arxiv.org/abs/2101.05320}{{\ttfamily 2101.05320}}.

\end{thebibliography}\endgroup

\end{document}